\documentclass[10pt,conference]{IEEEtran}
\IEEEoverridecommandlockouts
\usepackage{acro}[=v2]
\DeclareAcronym{RAN}{
    short=RAN,
    long=Radio Access Network
}

\DeclareAcronym{O-RAN}{
    short=O-RAN,
    long=Open \acl{RAN}
}

\DeclareAcronym{RIC}{
    short=RIC,
    long=\ac{RAN} Intelligent Controller
}

\DeclareAcronym{non-RT RIC}{
    short=Non-RT \ac{RIC},
    long=Non Real-time \ac{RIC}
}

\DeclareAcronym{near-RT RIC}{
    short=Near-RT \ac{RIC},
    long=Near Real-Time \ac{RIC}
}

\DeclareAcronym{RT RIC}{
    short=RT \ac{RIC},
    long=Real-Time \ac{RIC}
}

\DeclareAcronym{SMO}{
    short=SMO,
    long=Service Management and Orchestration
}

\DeclareAcronym{KPI}{
    short=KPI,
    long=Key Performance Indicator
}

\DeclareAcronym{ML}{
    short=ML,
    long=Machine Learning
}

\DeclareAcronym{MNO}{
    short=MNO,
    long=Mobile Network Operator
}

\DeclareAcronym{MCS}{
    short=MCS,
    long=Modulation and Coding Scheme
}

\DeclareAcronym{GraphSAGE}{
    short=GraphSAGE,
    long=Graph Sample and Aggregate
}

\DeclareAcronym{GNN}{
    short=GNN,
    long=Graph Neural Network
}

\DeclareAcronym{DQN}{
    short=DQN,
    long=Deep Q-Network
}

\DeclareAcronym{MSE}{
    short=MSE,
    long=Mean Squared Error
}

\DeclareAcronym{CU}{
    short=CU,
    long=Central Unit
}

\DeclareAcronym{DU}{
    short=DU,
    long=Distributed Unit
}

\DeclareAcronym{CIO}{
    short=CIO,
    long=Cell Individual Offset
}

\usepackage{cite}
\usepackage{amsmath,amssymb,amsfonts}
\usepackage{algorithm} %added
\usepackage{algorithmic}
\usepackage{graphicx}
\usepackage{subcaption}
\usepackage{textcomp}
\usepackage{units}
\usepackage{multicol}
\usepackage{fancyhdr}
\usepackage[table,xcdraw]{xcolor}
\usepackage[letterpaper,left=0.625in,right=0.625in,top=0.75in, bottom=1.04in]{geometry}

\iftrue
\usepackage{titlesec}

\titlespacing*{\section}{0pt}{\parskip}{-\parskip}
\titlespacing{\subsection}{0pt}{\parskip}{-\parskip}
\titlespacing{\subsubsection}{1em}{\parskip}{-\parskip}
\fi

\def\BibTeX{{\rm B\kern-.05em{\sc i\kern-.025em b}\kern-.08em
    T\kern-.1667em\lower.7ex\hbox{E}\kern-.125emX}}

\usepackage{soul} %nice highlights

\newcommand{\Gc}{\mathcal{G}}
\newcommand{\AAc}{\mathcal{A}}
\newcommand{\Pc}{\mathcal{P}}
\newcommand{\Kc}{\mathcal{K}}
\newcommand{\Vc}{\mathcal{V}}
\newcommand{\Ec}{\mathcal{E}}

\newcommand{\Af}{\mathbf{A}}

\begin{document}
% Control the number of names in references
\bstctlcite{IEEEexample:BSTcontrol}

\title{Learning and Reconstructing Conflicts in O-RAN: A Graph Neural Network Approach}

\author{
  \IEEEauthorblockN{
    Arshia Zolghadr,
    Joao F. Santos,
    Luiz A. DaSilva,
    Jacek Kibi\l{}da
  }
%  \vspace{0.75em}
  \IEEEauthorblockA{
    \textit{Commonwealth Cyber Initiative, Virginia Tech}, USA, e-mail: \{arshiaz,joaosantos,ldasilva,jkibilda\}@vt.edu
  }
}

\maketitle

\begin{abstract}
The \ac{O-RAN} architecture enables the deployment of third-party applications on the \acp{RIC}. 
However, the operation of third-party applications in the \ac{near-RT RIC}, known as xApps, may result in conflicting interactions. 
Each xApp can independently modify the same control parameters to achieve distinct outcomes, which has the potential to cause performance degradation and network instability. 
The current conflict detection and mitigation solutions in the literature assume that all conflicts are known a priori, which does not always hold due to complex and often hidden relationships between control parameters and \acp{KPI}. In this paper, we introduce the first data-driven method for reconstructing and labeling conflict graphs in \ac{O-RAN}. Specifically, we leverage GraphSAGE, an inductive learning framework, to dynamically learn the hidden relationships between xApps, parameters, and \acp{KPI}. Our numerical results, based on a conflict model used in the \ac{O-RAN} conflict management literature, demonstrate that our proposed method can effectively reconstruct conflict graphs and identify the conflicts defined by the O-RAN Alliance. 
\end{abstract}


% Reset acronyms after the title and abstract
\acresetall

\iftrue
\fancypagestyle{firstpage}
{
    \fancyhead[L]{This work has been submitted to the IEEE for possible
      publication.\\
      Copyright may be transferred without notice, after which this version may no longer be accessible.}
    \fancyhead[R]{}
    \pagenumbering{gobble}
}
\fi

\section{Introduction}\label{sec:intr}
\thispagestyle{firstpage}

    The \ac{O-RAN} Alliance introduced an open architecture that disaggregates the \ac{RAN} into different functional components orchestrated under a management framework capable of running custom third-party applications~\cite{Polese2022UnderstandingChallenges}. This management framework comprises two \acp{RIC} operating in different timescales. The \ac{non-RT RIC} is a component of the \ac{SMO} and is in charge of hosting third-party applications, known as rApps, that implement long-term tasks, with timescales greater than \unit[1000]{ms}, e.g., data analytics, AI/ML model training, and inference to optimize the \ac{RAN}~\cite{d2022orchestran}. In contrast, the \ac{near-RT RIC} is in charge of hosting third-party applications, known as xApps, that implement time-sensitive tasks, with timescales on the order of 10--\unit[1000]{ms}.
     
    In the \ac{near-RT RIC}, the xApps act as plugin-like extensions that can control base stations to enhance the capabilities of the \ac{RAN} and provide network operators with different functionality, e.g., load balancing, mobile handover, and RAN slicing~\cite{santos2024managingorannetworksxapp}. The xApps may operate independently from each other, which has the 
    potential for conflicting interactions, as different plugins may modify the same control parameters to achieve distinct outcomes. For example, a plugin may optimize parameters to maximize throughput, whereas another plugin aims to minimize energy consumption, leading to performance degradation and reduced network stability and reliability~\cite{delPrever2024PACIFISTA:RAN}.

    The \ac{O-RAN} specifications identify three types of conflict that can occur within the \ac{near-RT RIC}~\cite{Polese2022UnderstandingChallenges}: direct, implicit, and indirect. These conflicts can then be defined in terms of their level of interaction required and observability. \emph{Direct conflicts} arise when plugins simultaneously attempt to modify the same control parameter, which can be identified before their deployment. For instance, one xApp may aim to optimize throughput by increasing transmit power, while another xApp may attempt to minimize energy consumption by reducing it. \emph{Indirect conflicts} arise when xApps modify distinct parameters that inadvertently affect the same \ac{KPI}, which are less obvious and cannot always be antiicipated. For example, different xApps can modify the Cell Individual Offset (CIO) and antenna tilts, control parameters that affect a single \ac{KPI}: the cell's handover boundary. Finally, \emph{implicit conflicts} emerge when a parameter implicitly influences another by affecting \acp{KPI} used by plugins to trigger further actions on other control parameters, leading to complex relationships that are difficult to observe. For example, a plugin modifying the \ac{RAN} slice allocation to affect user throughput may unintentionally interfere with another plugin monitoring the user throughput to control handovers.    

\begin{figure}[t]
    \centering
    \includegraphics[width=.90\columnwidth]{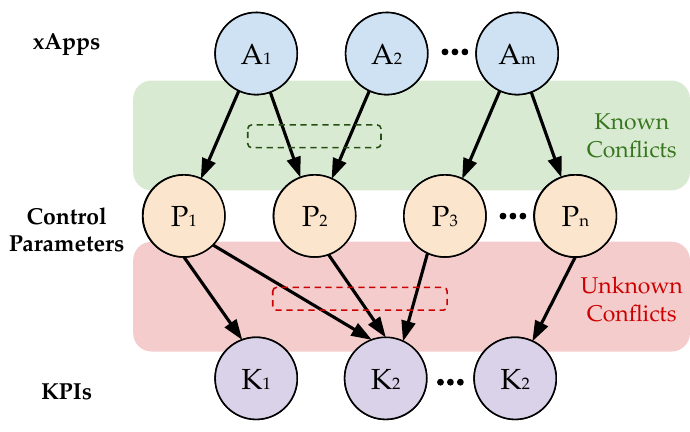}
    \vspace{-0.5em}
\caption{Conflict graph illustrating potential known conflicts between xApps and their control parameters (identifiable before deployment), and unknown conflicts that may arise due to the non-trivial relationships between control parameters and KPIs. 
}
    \label{fig:intro}
    \vspace{-2em}
\end{figure}

Despite growing industry and regulatory interest in \ac{O-RAN} deployments and the recent surge in related research efforts, few works~\cite{Wadud2023ConflictApproach,Zhang2022TeamO-RAN,Adamczyk2023ConflictRIC,delPrever2024PACIFISTA:RAN} explore conflict detection and mitigation. For instance, both~\cite{Wadud2023ConflictApproach} and~\cite{Zhang2022TeamO-RAN} focused on fundamental approaches to conflict mitigation, assuming that all relationships between xApps, control parameters, and \acp{KPI} are known a priori. The relationship between xApps and parameters is known by design, since xApps explicitly inform the Subscription Manager on the \ac{near-RT RIC} of their subscriptions to modify  control parameters~\cite{santos2024managingorannetworksxapp}.
However, the relationship between control parameters and \acp{KPI} is often non-trivial, dynamic, and scenario-dependent~\cite{Adamczyk2023ConflictRIC}, and such, indirect and implicit conflicts may be unknown in advance, as shown in Fig.~\ref{fig:intro}. 

To tackle this challenge,~\cite{delPrever2024PACIFISTA:RAN} and~\cite{Adamczyk2023ConflictRIC} proposed conflict detection solutions where conflicts are not necessarily known in advance. The work of~\cite{Adamczyk2023ConflictRIC} proposed an xApp prioritization strategy that manually groups control parameters affecting the same \acp{KPI}, and 
identifies conflicts within the group when a predefined performance degradation is observed. However, this approach is only suitable for the detection of direct and indirect conflicts. 
PACIFISTA~\cite{delPrever2024PACIFISTA:RAN} proposed modeling the relationships between xApps, control parameters, and \acp{KPI} as heterogenous conflict graphs. This approach enables the detection of conflicts by analyzing the structure of the graph to identify well-defined topologies representing different types of conflicts. To create conflict graphs, they adopt a profiling pipeline that tests and makes a statistical profile of xApps in a sandbox environment, similar to a digital twin. However, the effectiveness of their conflict graph (and consequently conflict detection) relies on the quality of an arbitrary set of tests and the fidelity of the sandbox to capture a realistic environment.

In this work, we build upon the conflict graph modeling approach introduced in~\cite{delPrever2024PACIFISTA:RAN}, addressing its reliance on arbitrary testing within digital twins. We propose a novel data-driven method using \acp{GNN} to learn the relationships between the nodes (e.g., xApps, control parameters, and KPIs) based on 
collected data from the current state of the xApps, parameters, and KPIs.
By leveraging \acp{GNN}, we can predict links and identify hidden relationships based on the graph topology and node features, allowing us to reconstruct complete conflict graphs. Specifically, we leverage GraphSAGE~\cite{Hamilton2017InductiveGraphs}, an inductive learning framework, for its ability to generalize to unseen nodes and scale efficiently across large graphs.
%In this work, we build on the modeling approach introduced in~\cite{delPrever2024PACIFISTA:RAN} that represents the relationships between xApps, control parameters, and \acp{KPI} as heterogeneous conflict graphs to overcome its reliance on arbitrary testing in a digital twin and instead learn the relationships and reconstruct conflict graphs based on data from RAN. 
%In this work, we build on the ideas presented in~\cite{delPrever2024PACIFISTA:RAN} and explicitly define the O-RAN near-RT control loop as a heterogeneous conflict graph, which must be reconstructed from the \ac{RAN} data.
%This leads us to propose a new data-driven method for conflict graph reconstruction based on \acp{GNN}. \acp{GNN} are known to excel at link prediction by leveraging graph topology and node features to identify hidden relationships. Specifically, we leverage \acs{GraphSAGE}, an inductive learning framework, for its ability to generalize to unseen nodes and scale efficiently across large graphs. The proposed method dynamically learns interactions between xApps, control parameters, and \acp{KPI} in a \ac{RAN} to uncover relationships and reconstruct conflict graphs, which encapsulate all types of conflicts present in the \ac{RAN} and can be identified using graph-theoretic conflict definitions adopted from~\cite{delPrever2024PACIFISTA:RAN}. 
In possession of the reconstructed conflict graph, we can utilize graph labeling to identify the different types of conflicts based on their well-defined, graph-based definition. This approach allows us to detect conflicts that emerge from the learned interactions between control parameters and \acp{KPI}.
    
    The main contributions of this paper are as follows:
    \begin{itemize}
        \item We propose a data-driven, \ac{GNN}-based method for reconstructing conflict graphs, capturing sequential interactions between the nodes of the conflict graph, e.g, where xApps modify parameters, parameters affect \acp{KPI}, and \acp{KPI} provide feedback to xApps.
        
        \item We propose graph-based definitions of conflicts, based on the definitions proposed in~\cite{delPrever2024PACIFISTA:RAN}, for identifying the three types of conflicts considered by the O-RAN Alliance. 
        
        \item We validate our conflict graph reconstruction and evaluate 
        our conflict labeling solutions using Gaussian distributed \acp{KPI} generated from a conflict model used in~\cite{Banerjee2022TowardNetworks,Wadud2023ConflictApproach}.
    \end{itemize}

To the best of our knowledge, this is the first work to propose a conflict graph reconstruction method for learning unknown relationships and conflicts in \ac{O-RAN}.
The remainder of this paper is organized as follows. 
In Section~\ref{sec:prob}, we pose our problem statement on the reconstruction of conflict graphs. 
In Section~\ref{sec:conf}, we detail our \ac{GNN}-based method to learn from complex relationships from collected data and reconstruct conflict graphs. 
In Section~\ref{sec:meth}, we describe graph-based definitions for the three types of conflicts in \ac{O-RAN}.
In Section~\ref{sec:eval}, we validate our solution numerically using a conflict model from the literature.
Finally, in Section~\ref{sec:conc}, we summarize our findings and discuss avenues for future work.

\section{Problem Statement}\label{sec:prob}

We define a heterogeneous graph $\Gc=(\Vc,\Ec)$ where $\Vc$ is a set of vertices and $\Ec$ is a set of edges. The topological structure of $\Gc$ can be represented by an adjacency matrix $\Af$. We assume that $\Gc$ is a vertex-labeled graph and that there are three categories of vertices:
\textit{(i)} plugins deployed in the \ac{near-RT RIC} (denoted as $\AAc$);
\textit{(ii)} control parameters that can be modified by plugins (denoted as $\Pc$); and \
\textit{(iii)} \ac{RAN} \acp{KPI} that can be affected by changes to parameters (denoted as $\Kc$).
Therefore, $\Vc = \AAc \cup \Pc \cup \Kc$.
In addition, an edge between $\AAc$, $\Pc$, and $\Kc$ can be interpreted as a valid relationship between the vertices. For instance, the existence of an edge between $\AAc$ and $\Pc$ or $\AAc$ and $\Kc$ represents the subscription of xApps to control parameters or monitor \acp{KPI}, respectively.
Similarly, an edge between $\Pc$ and $\Kc$ shows the influence of a parameter on a \ac{KPI}.
While some relationships in the heterogeneous graph representation of conflicts are known by design, i.e., the relationships between $\AAc$ and $\Pc$ and between $\AAc$ and $\Kc$ through the subscription process between xApps and the \ac{near-RT RIC}'s Subscription Manager~\cite{Polese2022UnderstandingChallenges}~\cite{santos2024managingorannetworksxapp}, others are less straightforward. Specifically, the relationships between $\Pc$ and $\Kc$ are often non-trivial, dynamic, and scenario-dependent, and must be inferred from the 
collected data from the current state of the xApps, parameters, and \acp{KPI}~\cite{delPrever2024PACIFISTA:RAN}. Therefore, our objective is to estimate the adjacency matrix $\Af$ based on the collected data.

\section{Conflict Graph Reconstruction in O-RAN}\label{sec:conf}

     We propose a \ac{GNN}-based approach for reconstructing conflict graphs based on collected data.
     To accomplish this, we leverage \acs{GraphSAGE}~\cite{Hamilton2017InductiveGraphs}, an inductive learning framework that learns vertex embeddings by aggregating and transforming feature information from neighboring vertices.
     Unlike conventional neural networks that process inputs independently, GraphSAGE requires a graph structure as input,
     making it particularly suitable for capturing complex relationships that evolve over time as the system progresses.
     To represent the data from the \ac{RAN}, we construct a multivariate time series temporal graph \(\mathcal{G}_T=(\mathcal{V}_T, \mathcal{E}_T)\), where each vertex \(v_t \in \mathcal{V}_T\) represents time step $t$ and is associated with a feature vector \(\mathbf{x}_t\), containing concatenated values of $\Pc$ and $\Kc$ observed at time step $t$.
     We assume a temporal dependency limited to consecutive vertices, i.e., that our features at time step $t$ only affect the values of features at $t+1$, and as such, we define $(v_t,v_{t+1}) \in \mathcal{E}_T$.
     This representation structure is inspired by temporal graph tasks, where time-stamped data is modeled as a graph to capture relationships across time steps~\cite{xu2023timegnn}.

    % In this structure, the temporal graph is the input graph to the neural network, and its vertices $v_t$ correspond to a node in the GraphSAGE neural network. The feature vector associated with each vertex serves as the input node attribute for the first layer of the neural network.The neighborhood \(N(\mathbf{v}_t)\)of vertex $v_t$ is defined by its sequential neighbors, the previous time step $v_{t-1}$ and the next time step $v{t+1}$
    The temporal graph serves as the structural backbone for the GraphSAGE neural network, where each vertex $v_t$ is mapped to a node, and its feature vector $\mathbf{x}_t$ becomes the input attribute for the first layer of the \ac{GNN}. Since we assume a temporal dependency limited to consecutive vertices, the neighborhood \(N(v_t)\) of vertex $v_t$ only includes the previous time step $v_{t-1}$ and the next time step $v_{t+1}$. The GraphSAGE framework processes the graph interactively in layers, where each layer updates the feature embedding of each vertex $v_t$ by aggregating information from its neighbors. At each layer \(k\), the feature embedding of vertex \(v_t\) denoted as \(\mathbf{h}_{v_t}^{(k)}\), is computed as following the approach in~\cite{Hamilton2017InductiveGraphs}:
    \vspace{-0.5em}
    \begin{equation}
     \mathbf{h}_{v_t}^{(k)} = \sigma \left( \mathbf{W}^k \cdot \mathbb{E} \left[ \{ \mathbf{h}_{v_t}^{(k-1)} \} \cup \{ \mathbf{h}_u^{(k-1)} \, : \, \forall u \in N(v_t) \} \right] \right),
    \end{equation}

    where \(\mathbf{W}^k\in \Re^{{d_{\text{out}}}\times{d_{\text{in}}}}\) is the learnable weight matrix at layer \(k\) with input dimension \(d_{\text{in}}\) and output dimension \(d_{\text{out}}\), \(\sigma\) is the non-linear activation function, \(\mathbb{E}\) represents the element-wise mean of the neighborhood \(N(v_t)\), and the $(\cdot)$ symbol represents the matrix multiplication operation.
        Consequently, the embeddings at the final layer, \(\mathbf{h}_{v}^{(K)}\), where \(K\) denotes the number of layers, represent the learned latent representations of the vertices.
    We then train the model using the \ac{MSE} loss to minimize the difference between the reconstructed and input embeddings. The loss function is defined as:
    %\vspace{-0.5em}
    \begin{equation}
    L_{\text{MSE}} = \frac{1}{|\mathcal{V}_T|} \sum_{t=1}^{T} \left\lVert \mathbf{h}_{v_t}^{(K)} - \mathbf{x}_t \right\rVert^2,
    \vspace{-0.5em}
    \end{equation}%\vspace{1em}
    where \(T\) is the set of all timestamps, \(\mathbf{x}_t\) is the input feature vector at time \(t\), \({|\mathcal{V}_T|}\) is the number of vertices, \(\left\lVert \right\rVert^2\) denotes the squared euclidean norm, and \(\mathbf{h}_{v_t}^{(K)}\) is the predicted embedding of vertex \(v_t\).

    % Once our training is complete, the final embedding from the last training epoch \(\mathbf{h}_{v_t}^*\) encode both the temporal relationships and feature dependencies. These embeddings are used to compute pairwise correlation across all features that describe the desired conflict graph. The resulting correlation matrix Specifically, we calculate the pairwise Pearson correlation coefficient for each feature pair and apply a fixed threshold to binarize the correlations and obtain a reconstructed adjacency matrix $\hat{\Af}$ ~\cite{Blyth1994KarlCurve}. An element $a_{ij}$ of $\hat{\Af}$ represents the existence of an interaction between two features corresponding to the $i$-th row and $j$-th column of the adjacency matrix.
    Once our training is complete, the final embedding from the last training epoch encodes both the temporal relationships and feature dependencies. These embeddings are used to compute pairwise correlation across all features that describe the desired conflict graph. Finally, we apply a fixed threshold to binarize the correlations and obtain a reconstructed adjacency matrix $\hat{\Af}$, representing the conflict graph~\cite{Blyth1994KarlCurve}. The fixed threshold is empirically chosen to balance sensitivity to weak correlations in the reconstructed conflict graph.
    % Therefore, element $a_{ij}$ of $\hat{\Af}$ represents the existence of an interaction between two features corresponding to the $i$-th row and $j$-th column of the adjacency matrix.
    In post-processing, we complete our reconstructed adjacency matrix by leveraging the known information about the subscription of xApps to parameters and \acp{KPI}, i.e., populating the matrix with existing edges between $\AAc$ and $\Pc$ as well as $\AAc$ and $\Kc$.

\begin{figure}
    \centering
    \hfill
    \begin{minipage}{0.4\columnwidth}
    \begin{subfigure}[t]{0.99\linewidth}
        \centering
        \includegraphics[width=\textwidth]{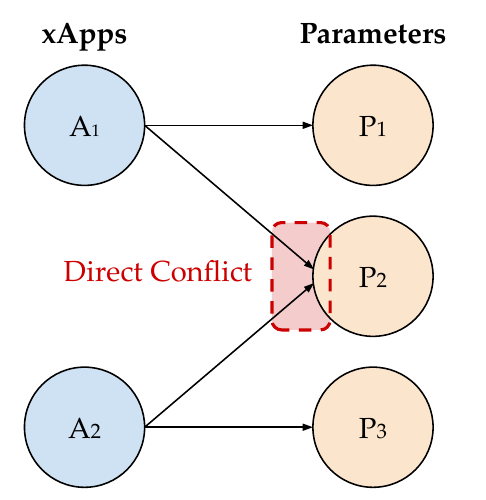}
        \caption{Direct conflict between xApps modifying the same control parameters.}
        \label{fig:dir_conf}
    \end{subfigure}

    \begin{subfigure}[b]{0.99\linewidth}
        \centering
        \includegraphics[width=\textwidth]{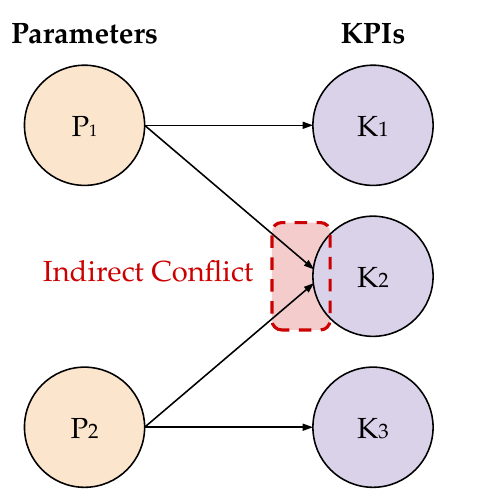}
        \caption{Indirect conflict between control parameters affecting the same \acp{KPI}.}
        \label{fig:indir_conf}
    \end{subfigure}
    \end{minipage}
    \hfill
    \begin{minipage}{0.405\columnwidth}
    \begin{subfigure}[b]{.99\linewidth}
        \centering
        \includegraphics[width=\textwidth]{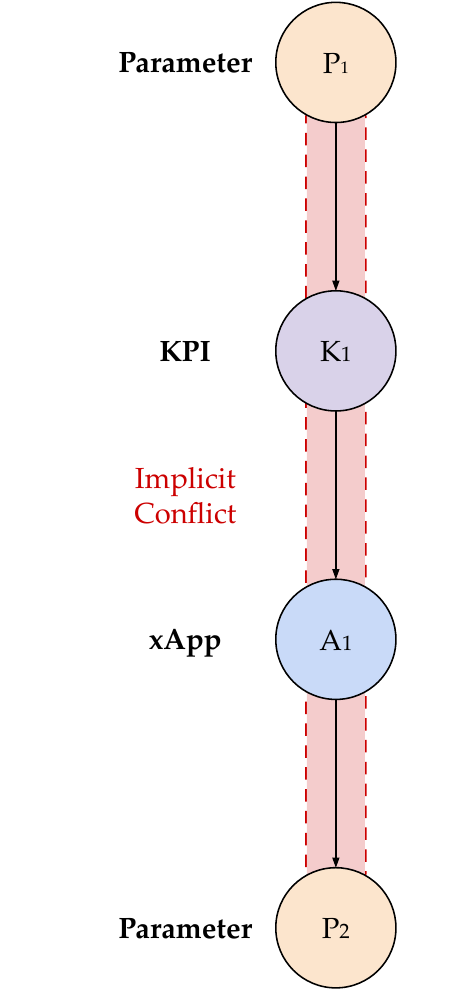}
        %\caption{Chains of relationships between parameters, \acp{KPI}, plugins, and other parameters causing implicit conflicts.}
        \caption{Chains of complex relationships create a logical dependency and cause an implicit conflict between different parameters.}
        \label{fig:imp_conf}
    \end{subfigure}
    \end{minipage}
    \hfill
    \caption{Illustration of the types of conflicts considered by the \ac{O-RAN} Alliance using our graph labeling definitions, allowing us to identify conflicts in the structure of a conflict graph $\Gc$. 
    %Illustration of the O-RAN conflict definitions.
    }
    \label{fig:diagconf}
    \vspace{-2.0em}
\end{figure}
\vspace{-0.25em}

\section{Labeling Conflicts in O-RAN}\label{sec:meth}

    %But since we now have a reconstructed adjacency $\Af$ matrix that includes that information, we can define a graph that uses this matrix as its topological structure.

    % These extend the definitions proposed in [4] by
        Based on the heterogeneous graph $\Gc$, we introduce definitions for the three types of conflicts considered by the O-RAN Alliance~\cite{O-RANWorkingGroup32023Near-RTArchitecture}.
        These extend the definitions proposed in~\cite{delPrever2024PACIFISTA:RAN} by
        %considering a causal influence of implicit relationships on the relation between parameters, as shown in Fig.~\ref{fig:diagconf}.
        diving deeper into the chains of complex relationships that give rise to indirect conflicts between different parameters, allowing us to understand their logical dependencies, as shown in Fig.~\ref{fig:diagconf}. This approach captures subtle and nuanced dependencies that might otherwise be overlooked, while also enabling the potential identification of the root causes of the indirect conflict.

    \begin{itemize}
        \item \textbf{Direct Conflicts}
        For any $a_i, a_j \in \AAc$, where $i\neq j$, \(a_i\) and \(a_j\) are in direct conflict if $e_{(a_i, p)}, e_{(a_j, p)} \in \Ec$. An example of two applications subscribing and attempting to modify the same parameter is shown in Fig.~\ref{fig:dir_conf}.

        \item \textbf{Indirect Conflicts}
        For any $a_i, a_j \in \AAc$ controlling $p_m, p_n\in \Pc$ respectively,
        and a \ac{KPI} $k\in \Kc$. \raisebox{0.1em}{\(a_i\)} and \raisebox{0.1em}{\(a_j\)} are in indirect conflict
        if $e_{(a_i, p_m)}, e_{(a_j, p_n)} ,e_{(p_m, k)},e_{(p_n, k)} \in \Ec$.
        An example of two parameters affecting the same \ac{KPI} \(k_2\) is
        shown in Fig.~\ref{fig:indir_conf}.

        \item \textbf{Implicit Conflicts}
        Considering any $a_i, a_j \in \AAc$ modifying $p_m, p_n\in \Pc$, respectively, and a \ac{KPI} $k\in \Kc$. \(a_i\) and \(a_j\) are in implicit conflict  if $e_{(a_i, p_m)}, e_{(p_m,k)}, e_{(k, a_j)}, e_{(a_j, p_n)} \in \Ec$. This type of conflict cannot be observed directly, and even the relationship between the xApps and parameters may not be obvious, as shown in Fig.~\ref{fig:imp_conf}.
    \end{itemize}

    \textbf{Remarks}:
    \begin{itemize}
        \item
        The definition of an implicit conflict implies a depth of conflicts, whereby the existence of an implicit conflict between $a_i$ and $a_j$, and $a_j$ and $a_k$ implies that there is an implicit conflict also between $a_i$ and $a_k$. In this preliminary work, we limit ourselves to identifying the depth-one implicit conflicts only.
    \item The existence of an implicit conflict between $a_i$ and $a_j$ (or $p_m$ and $p_n$) implies a dependency between $a_i$ and $a_j$ (or $p_m$ and $p_n$), which can be interpreted as an indirect conflict as assumed in~\cite{delPrever2024PACIFISTA:RAN}. Notwithstanding the merits of alternative conflict labeling approaches, we adhere to the conflict labels defined by the \ac{O-RAN} specifications~\cite{O-RANWorkingGroup32023Near-RTArchitecture}.

\end{itemize}

\section{Numerical Evaluation}\label{sec:eval}
In this section, we validate our \ac{GNN}-based method for reconstructing conflict graphs and detecting conflicts. First, we describe the conflict model and dataset used to train our \ac{GNN}, and the metric used in our evaluations.
Then, we assess the accuracy of our \ac{GNN} to reconstruct conflict graphs and our graph labeling solution to detect different types of conflicts.

\begin{figure}[t]
    \centering
    \begin{subfigure}[b]{0.49\columnwidth}
        \centering
        \includegraphics[width=\textwidth]{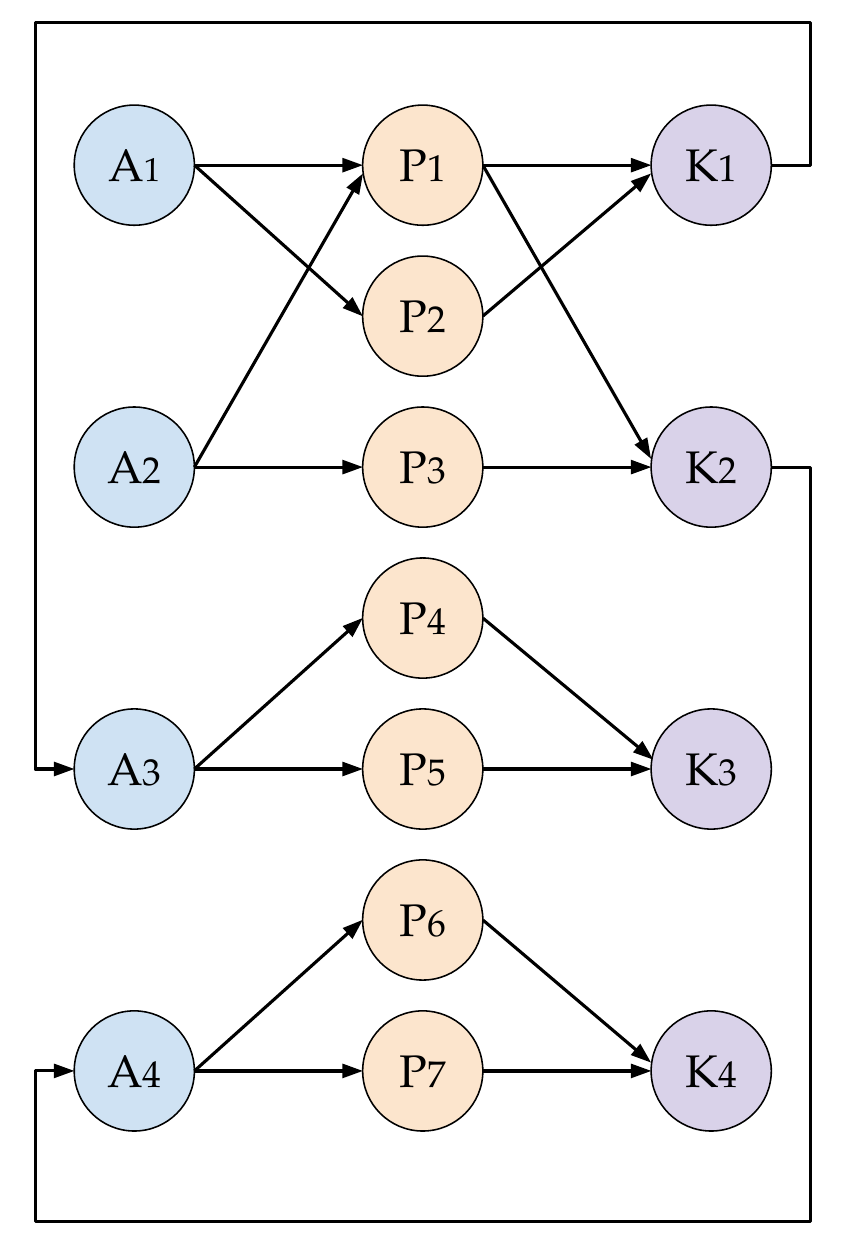}
        \caption{Exemplar conflict graph we adopt in our numerical evaluation.}
    \end{subfigure}
    \hfill
    \begin{subfigure}[b]{0.49\columnwidth}
        \centering
        \includegraphics[width=\textwidth]{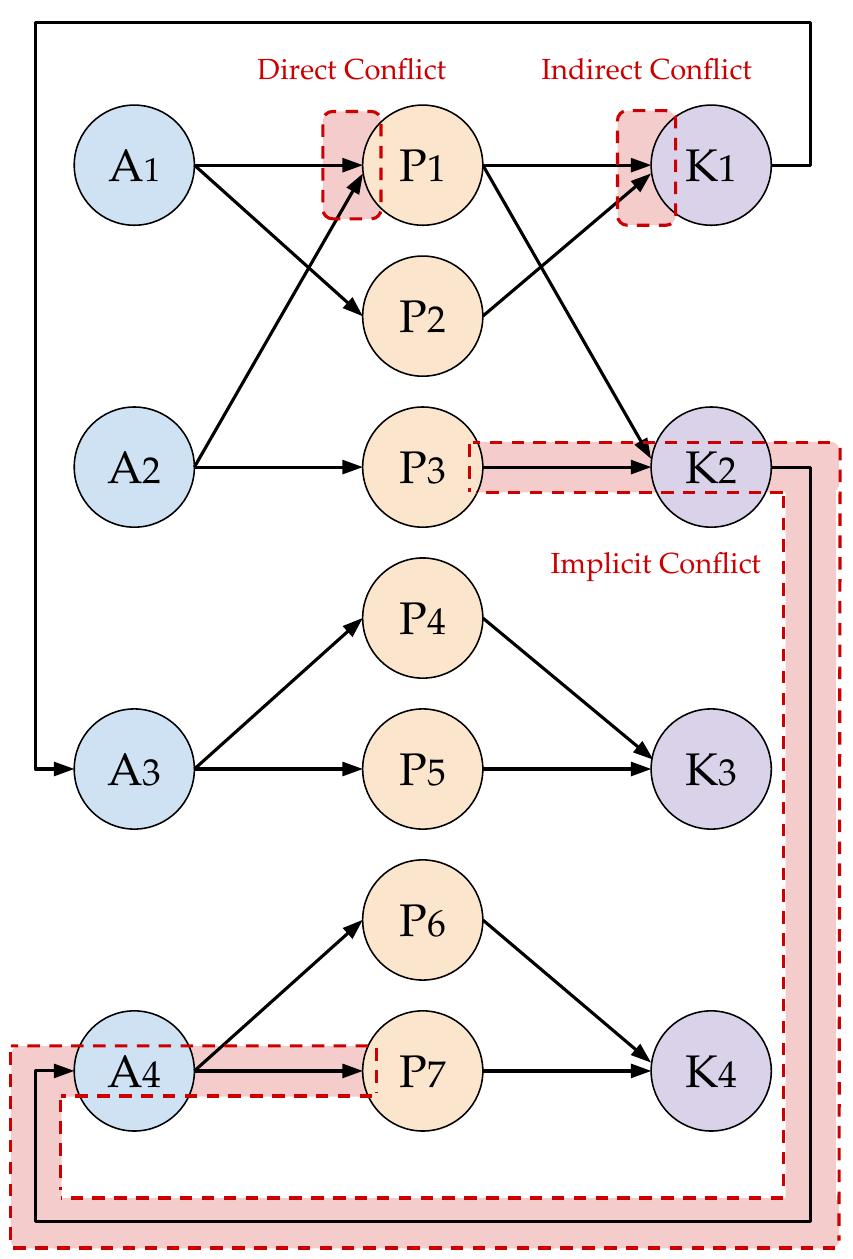}
        \caption{Example of conflict labeling according to our definitions.}
    \end{subfigure}
    \caption{Structure of the conflict model~\cite{Banerjee2022TowardNetworks} we used to validate our solutions for graph reconstruction and conflict labeling.}
    \label{fig:toy_model}
    \vspace{-2em}
\end{figure}

\subsection{Dataset, Network Architecture, and Evaluation Metric}

To evaluate our proposed \ac{GNN}-based method for learning the relationships between xApps, control metrics, and \acp{KPI}, and reconstructing conflict graphs, we collected a dataset based on the conflict model for cognitive autonomous networks proposed in~\cite{Banerjee2022TowardNetworks} and later adopted in other works on conflict management in \ac{O-RAN}~\cite{Wadud2023ConflictApproach}.
Fig.~\ref{fig:toy_model} shows a graphical representation of the conflict model and an example of the conflict labeling based on the definitions from Section~\ref{sec:meth}. The conflict model shows the interaction between four xApps that control seven parameters, which, in turn, influence four \acp{KPI}. We assume the \acp{KPI} follow a Gaussian distribution, with values defined by equations that describe the relationships between control parameters and \acp{KPI}, as follows: \(K_1 = 0.5 \cdot \exp\left(-\frac{(P_1 + 50)^2}{(2P_2)^2}\right),
K_2 = \exp\left(-\frac{(P_1 - 50)^2}{(2P_3)^2}\right), K_3 = \exp\left(-\frac{(P_4 + K_1)^2}{(2P_5)^2}\right), K_4 = \exp\left(-\frac{(P_7 + K_2)^2}{(2P_6)^2}\right).\)
Based on this model, we sampled control parameters using uniformly distributed random values within the ranges specified in Table~\ref{tab:pram} and calculated their effects on the corresponding \ac{KPI} to generate datasets (of arbitrary size) that captured their correlations.

To capture the unknown relationships in the conflict model, we designed a \ac{GNN} with 11 features, representing its seven parameters and four \acp{KPI}. The neural network architecture consists of three layers, where the input and output layers have sizes equal to the size of the dataset, and the hidden layer has 16 neurons. We also used a learning rate of 0.001.

\begin{table}[t]
\centering
\resizebox{\columnwidth}{!}{%
\begin{tabular}{cccccccc}
\rowcolor[HTML]{9FC5E8}
\textbf{Parameter} & $P_1$         & $P_2$         & $P_3$       & $P_4$       & $P_5$       & $P_6$       & $P_7$       \\
\rowcolor[HTML]{CFE2F3}
\textbf{Value Range}     & {[}0, 300{]} & {[}0, 300{]} & {[}0, 3{]} & {[}0, 3{]} & {[}0, 3{]} & {[}0, 3{]} & {[}0, 3{]}
\end{tabular}
}
\caption{Ranges of values used for parameters in our dataset.}
\label{tab:pram}
\vspace{-2em}
\end{table}

To evaluate our reconstruction and labeling, we adopt the F1 Score, the harmonic mean of precision and recall, as shown in Eq.~\ref{eq:f1}. This metric provides a balanced measure of accuracy due to its capacity to capture both false positives and negatives.

\small
\begin{equation}
\text{Precision} = \frac{\text{True Positive}}{\text{True Positive} + \text{False Positive}},
\end{equation}

\begin{equation}
\text{Recall} = \frac{\text{True Positive}}{\text{True Positive} + \text{False Negative}},
\end{equation}

\begin{equation}
\text{F1 Score} = 2 \cdot \frac{\text{Precision} \cdot \text{Recall}}{\text{Precision} + \text{Recall}}.
\label{eq:f1}
\end{equation}
\normalsize

\subsection{Graph Reconstruction Accuracy}

\begin{figure}[t]
    \centering
    \begin{subfigure}[b]{0.49\columnwidth}
        \centering
        \includegraphics[width=\textwidth]{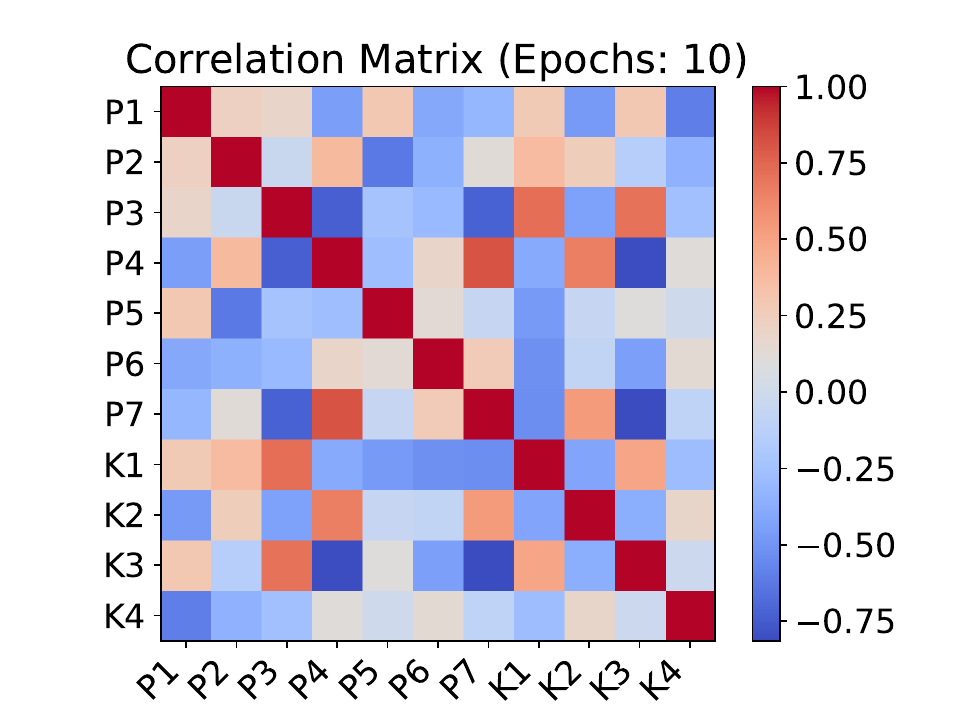}
        \vspace{-1em}
        \caption{Early stages of the correlation matrix at 10 epochs, showing the results of limited learning time that results in several false positives from spurious relationships.}
    \end{subfigure}
    \hfill
    \begin{subfigure}[b]{0.49\columnwidth}
        \centering
        \includegraphics[width=\textwidth]{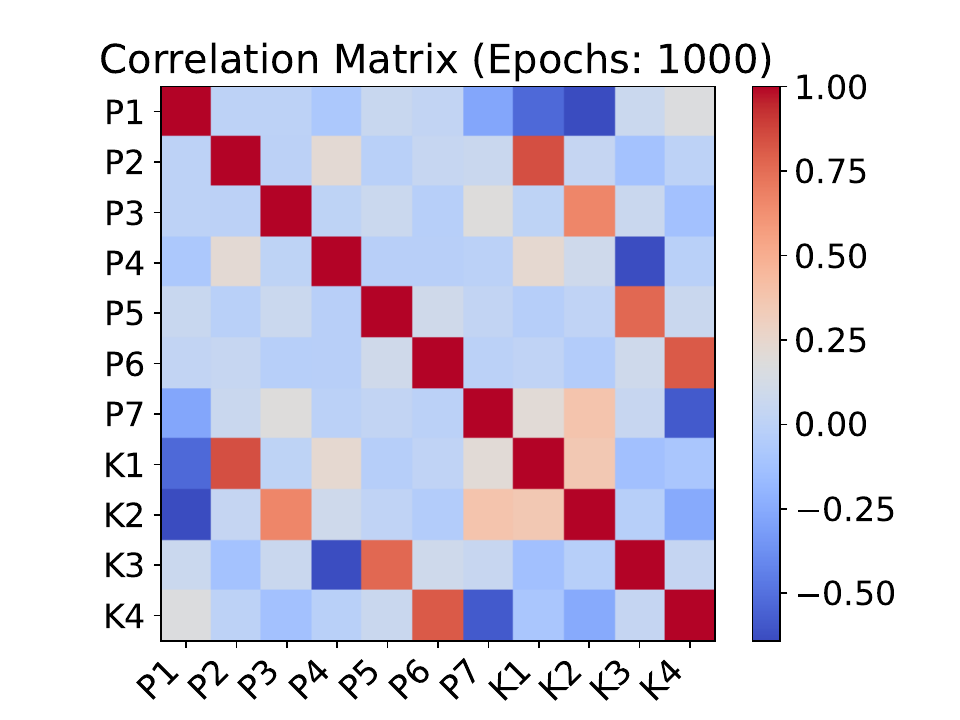}
        \vspace{-1em}
        \caption{Correlation matrix with longer training time at 1000 epochs, showing the model's convergence to learn the correct relationships between parameters and \acp{KPI}.}
    \end{subfigure}
    \caption{Evolution of the correlation matrix of our \ac{GNN} model with the number of epochs (created with a dataset size of 450 samples), illustrating the convergence.}
    \label{fig:visual}
    \vspace{-2em}
\end{figure}

%Motivate these results
In analysis, we are interested in assessing our \ac{GNN}'s ability to learn the relationships and accurately reconstruct conflict graphs. The performance of a \ac{GNN} model depends both on the amount of training information, i.e., the dataset size, and the training time, as the number of epochs allows the \ac{GNN} model to iteratively update its weights and better capture the underlying patterns from the graph. Fig.~\ref{fig:visual} presents the output correlation matrix generated by our \ac{GNN} model for different numbers of epochs, showing the benefits of additional training. However, some spurious edges with low correlation values persist even with additional epochs. While further training could reduce these edges, there is a risk of overfitting the \ac{GNN} to the dataset. To address this, we introduce a cutoff threshold for the correlation values, excluding edges below this threshold from the reconstructed conflict graph.

\begin{figure}[t]
\centering
\includegraphics[width=0.99\linewidth]{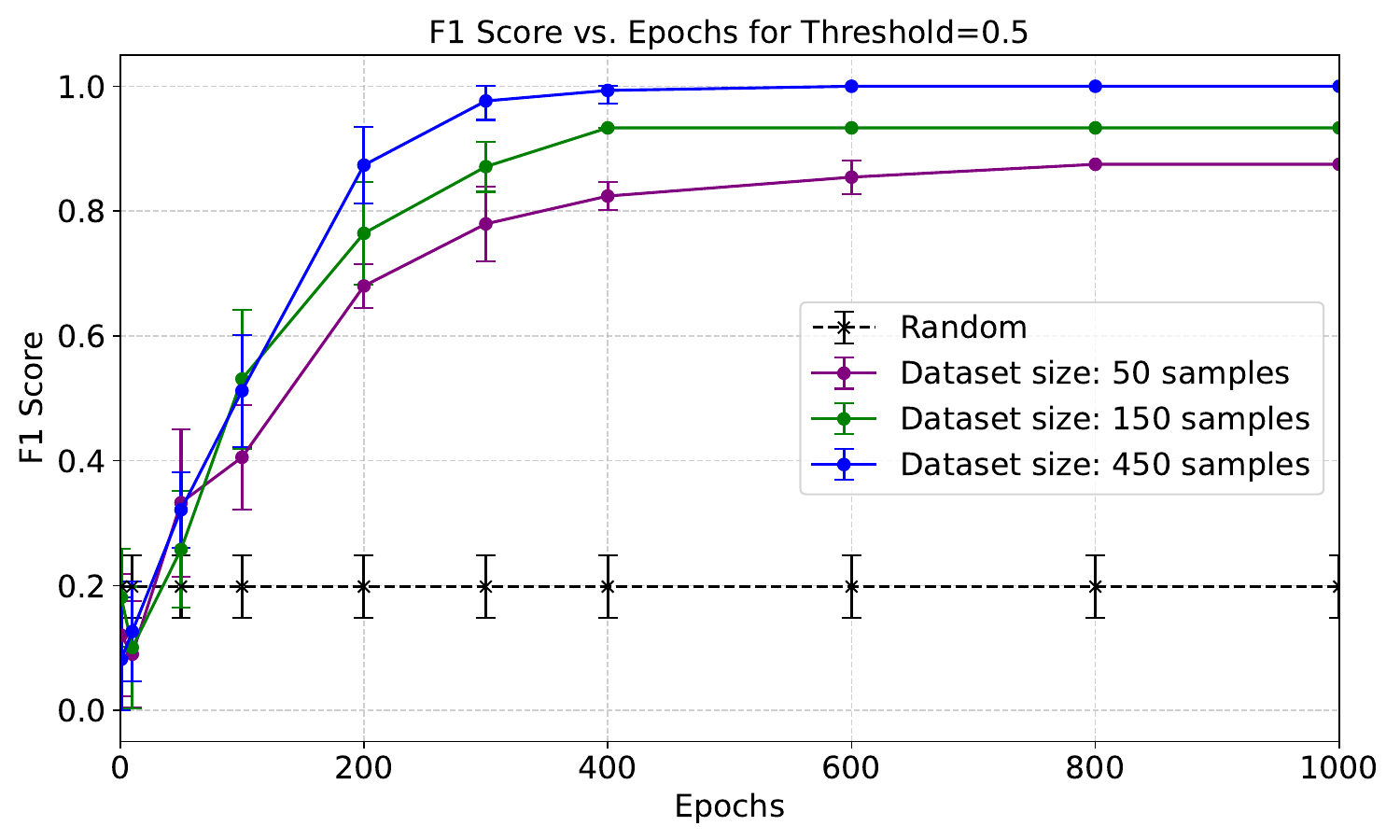}
\vspace{-2em}
\caption{Conflict graph reconstruction accuracy according to the number of epochs and dataset size for a\;fixed\;threshold\;of\;0.5.
}
\label{fig:f1_reconstruct}
\vspace{-1.5em}
\end{figure}

\begin{figure}[t]
\centering
\includegraphics[width=0.99\linewidth]{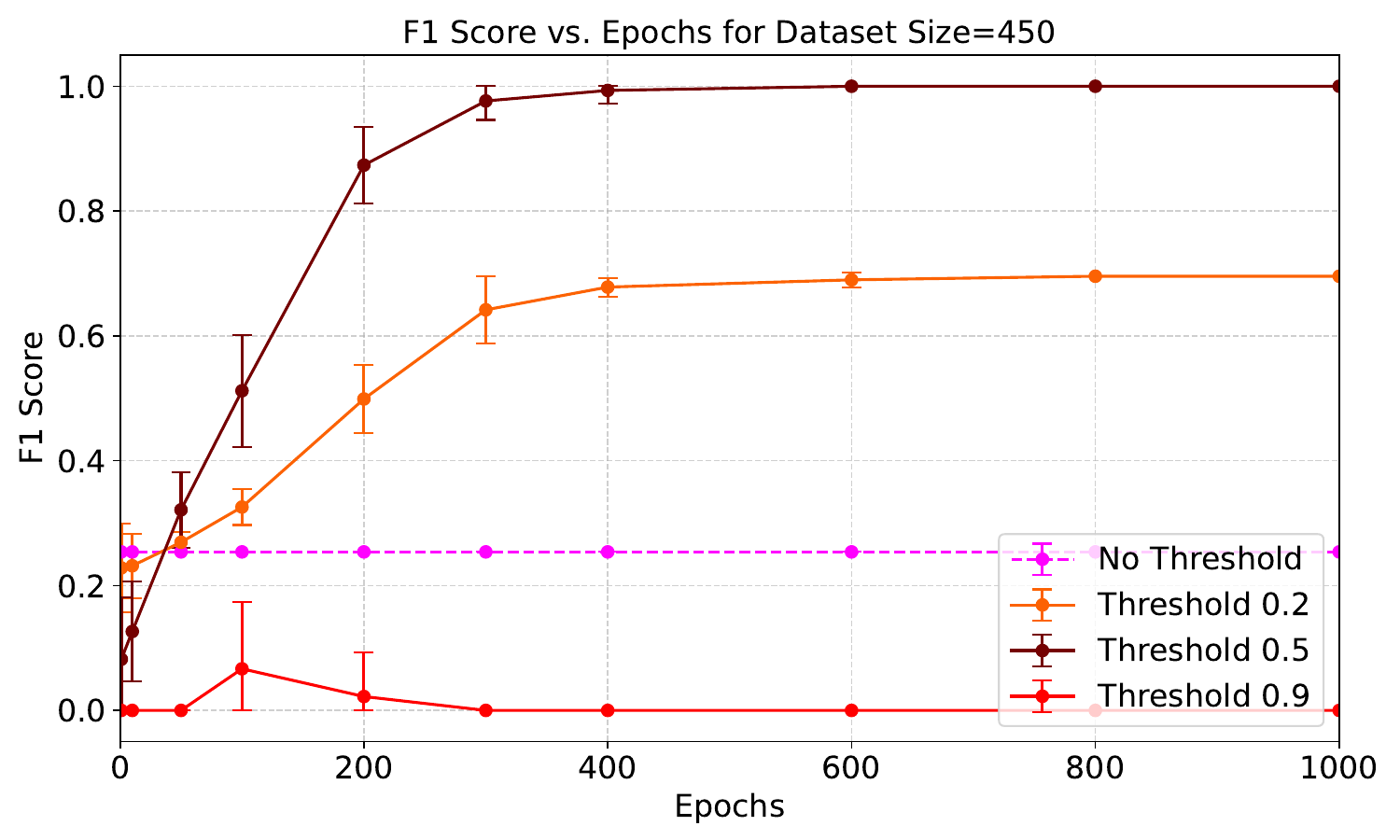}
\vspace{-2em}
\caption{Conflict graph reconstruction accuracy according to the epochs and
  thresholds for a fixed dataset size of 450 samples.
}
\label{fig:f1_reconstruct_threshold}
\vspace{-1.5em}
\end{figure}

In Fig.~\ref{fig:f1_reconstruct}, we evaluate our \ac{GNN} model reconstruction
accuracy using the F1 Score for different dataset sizes and a number of epochs,
with a fixed threshold of 0.5. To establish a baseline for comparison, we
evaluate our \ac{GNN}-based method for graph reconstruction against a uniformly
distributed random graph. We can observe the training time significantly influences the accuracy of the reconstructed graph up to 400 epochs, while the dataset size displays a bigger contribution to the accuracy of the model beyond this point. It is worth mentioning that we observed no substantial accuracy improvement with larger dataset sizes. These results show that our \ac{GNN}  model can successfully learn the relationships and recreate conflict graphs, achieving an accuracy of 100\% with at least 450 samples and 600  epochs.
In Fig.~\ref{fig:f1_reconstruct_threshold}, we evaluate the reconstruction accuracy of our \ac{GNN} for different thresholds and numbers of epochs, with a fixed dataset size of 450 samples. We observe that training time significantly impacts the accuracy of the reconstructed graph up to 400 epochs, while the threshold becomes more significant beyond this point.
These results demonstrate that incorporating a threshold improves accuracy by
eliminating false positives caused by edges
with low correlation values, achieving an accuracy of 100\% with a threshold of
0.5 and 600 epochs.

\subsection{Graph Labeling Accuracy}

In this analysis, we are interested in assessing the accuracy of our conflict labeling to identify different types of conflicts on a reconstructed conflict graph. The performance of the conflict labeling depends on the quality of the reconstructed conflict graph to capture the correct structure of the conflict model. Thus, we evaluate the conflict labeling based on the reconstructed graphs from the previous section. We only display results evaluating the identification of implicit and indirect conflicts, as we consistently obtained 100\% accuracy for direct conflicts across all settings, and the direct conflicts can be trivially identified through the analysis of the subscription information on the \ac{near-RT RIC}'s Subscription Manager~\cite{santos2024managingorannetworksxapp}.

\begin{figure}[t]
\centering
\includegraphics[width=0.99\linewidth]{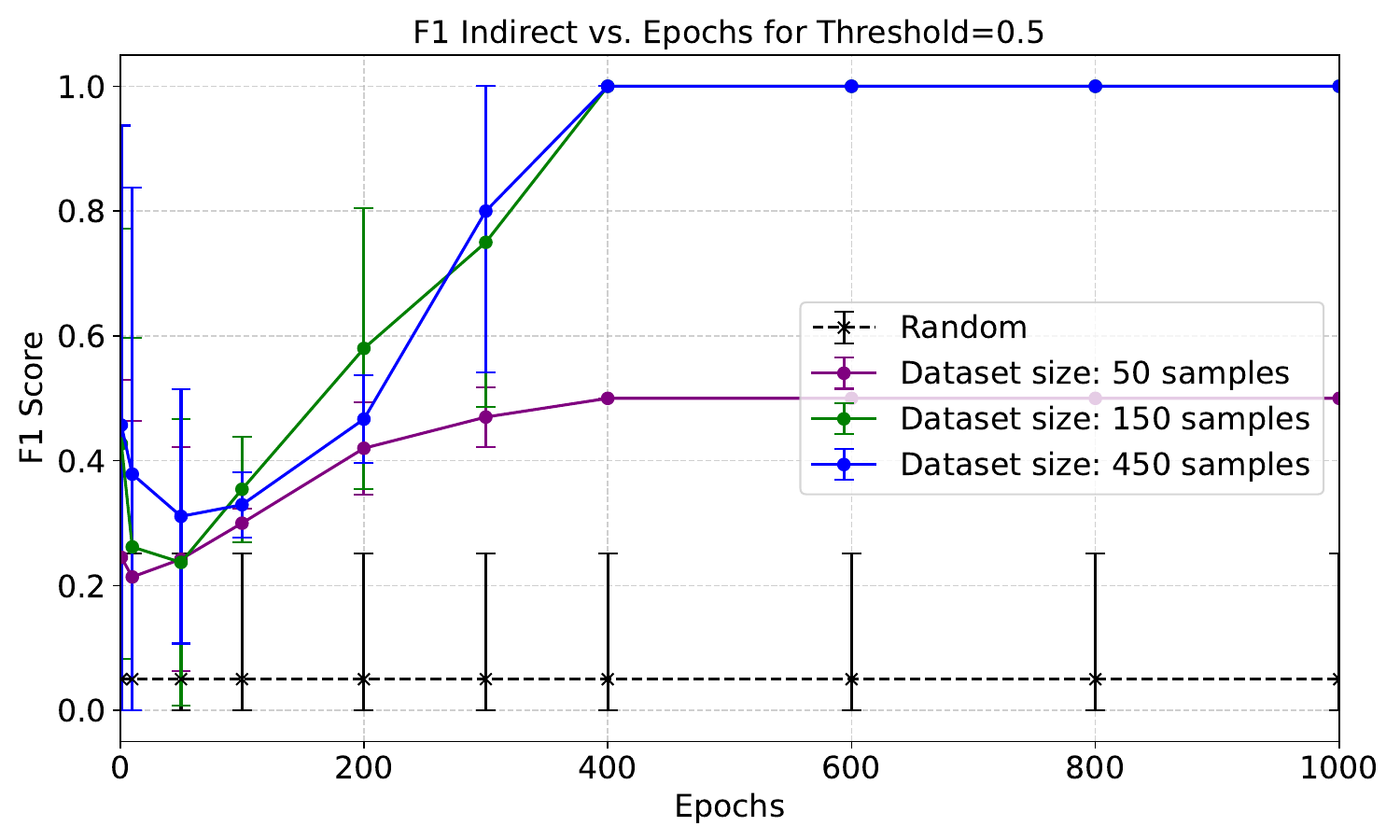}
\vspace{-2em}
\caption{Indirect conflict labeling accuracy according to the number of epochs
  and dataset size for a\;fixed\;threshold\;of\;0.5.
}
\label{fig:f1_indirect}
\vspace{-1.7em}
\end{figure}

\begin{figure}[t]
\centering
\includegraphics[width=0.99\linewidth]{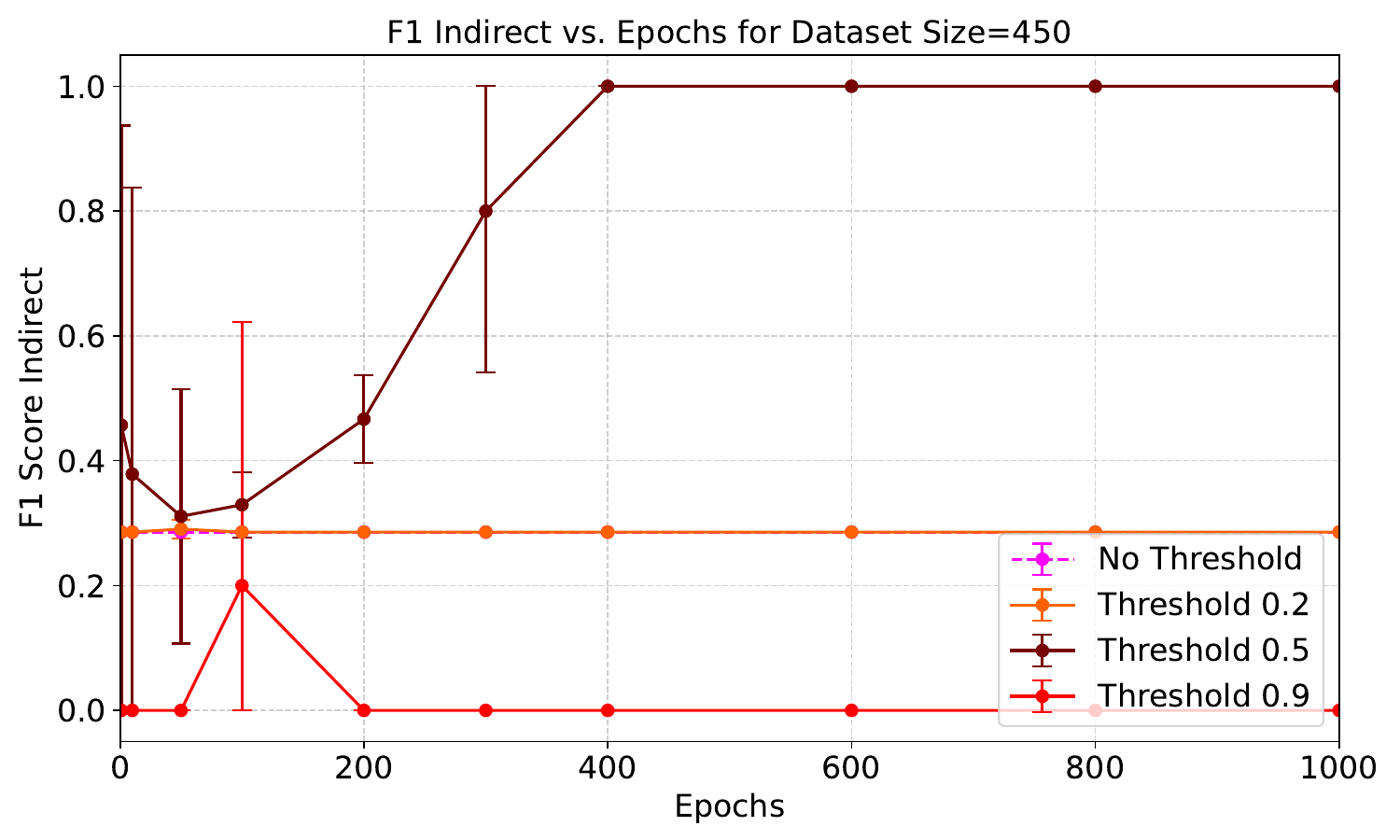}
\vspace{-2em}
\caption{Indirect conflict labeling accuracy according to the epochs and
  thresholds for a fixed dataset size of 450 samples.
}
\label{fig:f1_indirect_threshold}
\vspace{-1.5em}
\end{figure}

In Fig.~\ref{fig:f1_indirect}, we evaluate our graph labeling to detect indirect conflicts on reconstructed conflict graphs created using different dataset sizes and a number of epochs, with a fixed threshold of 0.5.
We observe that detecting indirect conflicts is considerably more challenging due to the effect of spurious edges on the models' accuracy. This complexity requires conflict graphs with higher reconstruction accuracy to capture correct edges without spurious edges giving false positive accuracy. Thus, accurately identifying indirect conflicts requires reconstructing conflict graphs with significantly longer training times, at least 600 epochs.
In Fig.~\ref{fig:f1_indirect_threshold}, we evaluate our graph labeling to detect indirect conflicts on reconstructed conflict graphs created using different thresholds and numbers of epochs, with a fixed dataset size of 450 samples. We observe that the threshold plays a crucial role in improving the accuracy of implicit conflict detection.
Due to the non-linear relationships between parameters and \acp{KPI} in our example model, many variables exhibit low correlations, leading to numerous spurious edges that result in false positives.
Our results show that a relatively high threshold of 0.5 offers a good balance to filter out edges with low correlation values to prune false positives and keep meaningful edges to prevent false negatives.

\begin{figure}[t]
\centering
\includegraphics[width=0.99\linewidth]{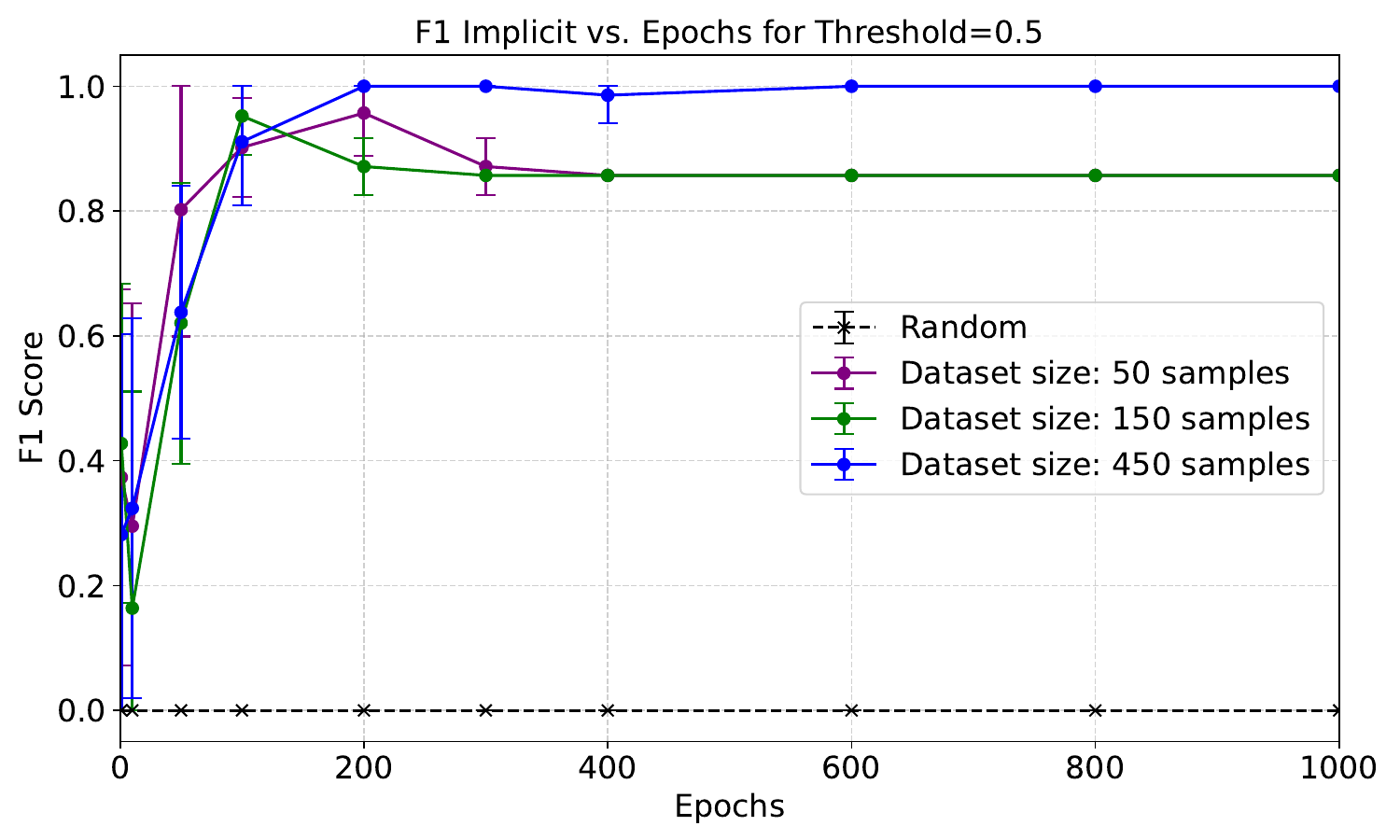}
\vspace{-2em}
\caption{
Implicit conflict labeling accuracy according to the number of epochs and dataset size for a\;fixed\;threshold\;of\;0.5.
}
\label{fig:f1_implicit}
\vspace{-1.5em}
\end{figure}

\begin{figure}[t]
\centering
\includegraphics[width=0.99\linewidth]{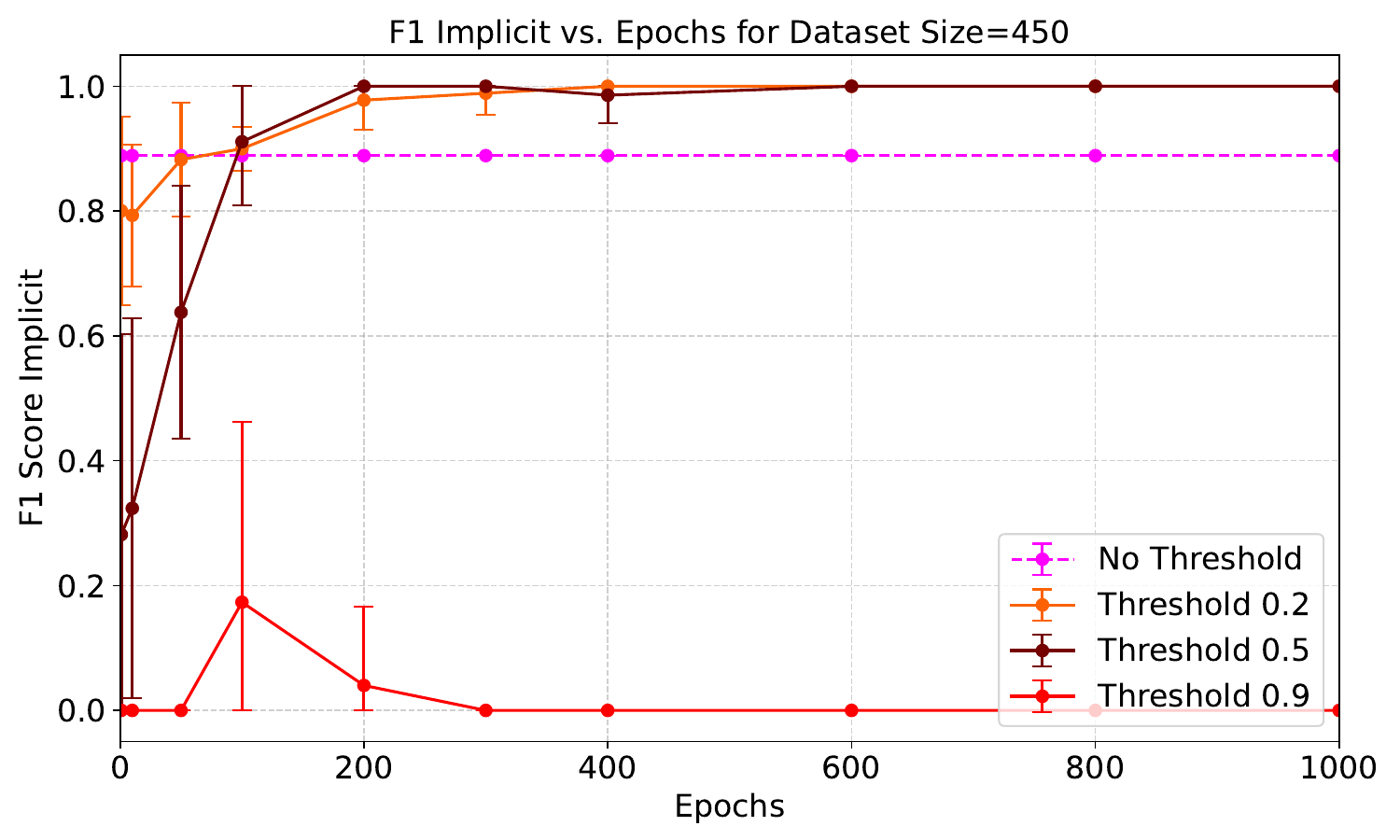}
\vspace{-2em}
\caption{
Implicit conflict labeling accuracy according to epochs and threshold values for a fixed dataset size of 450 samples.
}
\label{fig:f1_implicit_threshold}
\vspace{-2.0em}
\end{figure}

In Fig.~\ref{fig:f1_implicit}, we evaluate our graph labeling to detect implicit conflicts on reconstructed conflict graphs created using different dataset sizes and number of epochs, with a fixed threshold of 0.5.
As expected, the labeling performance improves with the accuracy of the reconstructed graph until 200 epochs (88\% reconstruction accuracy), and from there onward, conflict graphs generated with more data points display better performance.
In Fig.~\ref{fig:f1_implicit_threshold}, we evaluate our graph labeling to detect implicit conflicts on reconstructed conflict graphs created using different thresholds and numbers of epochs, with a fixed dataset size of 450 samples.
While converging akin to the previous plot, i.e., stabilizing after 300 epochs, we note that the threshold has only a minor impact on the labeling accuracy of implicit conflicts, provided it is not set very high, which can result in false negatives.
%The detection of implicit conflicts is notably high, often exceeding the accuracy of the reconstructed graph, as only a subset of correctly reconstructed edges is needed to identify these conflicts.
Our results show that we can achieve a 100\% detection of implicit conflicts when the graph is reconstructed with at least 450 samples, 200 epochs, and a threshold of 0.5.

\section{Conclusion}\label{sec:conc}

In this paper, we proposed the first data-driven \ac{GNN}-based method for reconstructing conflict graphs in \ac{O-RAN}. Our approach is designed to learn the relationships between xApps, control parameters, and \acp{KPI} to reconstruct complete conflict graphs that capture the three types of conflicts considered by the \ac{O-RAN} Alliance. In addition, we introduce a conflict labeling solution based on graph-based definitions for direct, indirect, and implicit conflicts. We validated our conflict graph reconstruction and evaluated our conflict labeling using a dataset generated from a known conflict model adopted in other works for conflict management in O-RAN, and demonstrated its accuracy across different dataset sizes, thresholds, and training epochs. Our results highlighted our ability to achieve high accuracy in graph reconstruction and conflict detection, particularly for implicit and indirect conflicts. It is worth mentioning that data-driven approaches such as ours may yield false negatives in cases where an xApp subscribes to a \ac{KPI} but does not actively use it to modify parameters, which might be a limitation in certain operational scenarios.
% In future works, we plan to extend our \ac{GNN} formulation to reconstruct directed graphs and consider causal relationships between xApps, control parameters, and \acp{KPI}, as a more effective approach to mitigating spurious edges and decreasing our dependency on correlation. In addition, we plan to explore the scalability of our solution for different numbers of xApps, control parameters, and \acp{KPI}, possibly using empirical data from real-world experiments. Finally, we aim to complement our work by developing conflict mitigation techniques to create a complete conflict management system for \ac{O-RAN}.
In future works, we plan to extend our \ac{GNN} formulation to reconstruct causal relationships between xApps, parameters, and \acp{KPI}, as a more effective approach to mitigating spurious edges and decreasing our dependency on correlation. In addition, we plan to explore the scalability of our solution for different numbers of xApps, parameters, and \acp{KPI}, leveraging empirical data obtained in an experimental setting. Finally, we aim to complement our work by developing conflict mitigation techniques to create a complete conflict management system for \ac{O-RAN}.

\section*{Acknowledgments}
The research leading to this paper received support\;from the Commonwealth Cyber Initiative, an investment in\;the advancement of\;cyber R\&D, innovation, and workforce development. For more information, visit: www.cyberinitiative.org.
This work also received support from the Horizon Europe SNS JU program
under grant No. 101139194 (6G-XCEL),
and from the National Science Foundation
 US-Ireland R\&D Partnership program under grant No. 2421362.

\bibliographystyle{IEEEtran}
\bibliography{IEEEabrv,revised_bib}

%\vspace{12pt}

\end{document}